\documentclass[12pt,preprint]{aastex}          
   \usepackage{graphicx}
   \usepackage{amssymb}
   \usepackage{epstopdf}

\def\teq#1{$\, #1\,$}
{\catcode`\@=11                                                  
   \gdef\SchlangeUnter#1#2{\lower2pt\vbox{\baselineskip 0pt\lineskip0pt    
   \ialign{$\m@th#1\hfil##\hfil$\crcr#2\crcr\sim\crcr}}}}           
\def\gtrsim{\mathrel{\mathpalette\SchlangeUnter>}}               
\def\lesssim{\mathrel{\mathpalette\SchlangeUnter<}}
             \font\sevenrm=cmr7

\def\pr{Phys. Rev.}                             
\def\app{Astroparticle Phys.}                   
\def\apss{Astr. Space Sci.}                     
\def\asr{Adv. Space Res.}                       
\def\prl{Phys. Rev. Lett.}                      
\def\ssr{Space Sci. Rev.}                       

\def\ThetaBnone{\Theta_{\hbox{\sevenrm Bn,u}}} 
   
\def\Machalf{{\cal M}_{\hbox{\sevenrm A}}}
\def\valf{v_{\hbox{\sevenrm A}}}
\def\Ualf{U_{\hbox{\sevenrm A}}}
\def\Chandra{{\it Chandra}}

\begin{document}
\newcommand{\vol}[2]{$\,$\rm #1\rm , #2.}                 
\newcommand{\figureoutpdf}[5]{\centerline{}
   \centerline{\hspace{#3in} \includegraphics[width=#2truein]{#1}}
   \vspace{#4truein} \caption{#5} \centerline{} }
\newcommand{\twofigureoutpdf}[3]{\centerline{}
   \centerline{\includegraphics[width=2.9truein]{#1}
        \hspace{0.4truein} \includegraphics[width=2.9truein]{#2}}
        \vspace{-0.25truein}
    \caption{{\small #3}} \vspace{-0.15truein}}    

\title{Topical Issues for Particle Acceleration Mechanisms in Astrophysical Shocks}

   \author{Matthew G. Baring}
   \affil{Department of Physics and Astronomy MS-108, Rice University, \\
      P.O. Box 1892, Houston, TX 77251, U.S.A. \it baring@rice.edu\rm}

\slugcomment{Accepted for publication in
             \it Astrophysics and Space Science\rm}

\begin{abstract}  
Particle acceleration at plasma shocks appears to be ubiquitous in the
universe, spanning systems in the heliosphere, supernova remnants, and
relativistic jets in distant active galaxies and gamma-ray bursts.  This
review addresses some of the key issues for shock acceleration theory
that require resolution in order to propel our understanding of particle
energization in astrophysical environments. These include magnetic field
amplification in shock ramps, the non-linear hydrodynamic interplay
between thermal ions and their extremely energetic counterparts
possessing ultrarelativistic energies, and the ability to inject and
accelerate electrons in both non-relativistic and relativistic shocks. 
Recent observational developments that impact these issues are
summarized.  While these topics are currently being probed by
astrophysicists using numerical simulations, they are also ripe for
investigation in laboratory experiments, which potentially can provide
valuable insights into the physics of cosmic shocks.
\vspace{-20pt}
\end{abstract}  

\section{Introduction}
 \label{sec:intro}

Supersonic flows abound in the cosmos, as do emission regions exhibiting
non-thermal radiation.  The intimate connection between the two
establishes that particle acceleration in astrophysical shocks is
germane to many systems, ranging from the heliosphere, to stars of
various sorts expelling winds, to supernova remnants, to extragalactic
jets and gamma-ray bursts. In the case of heliospheric shocks such as
travelling interplanetary discontinuities and planetary bow shock
environs, we can immerse ourselves in the plasma experiment via {\it in
situ} spacecraft measurements of non-thermal ions, electrons and
turbulent magnetic fields.  While localized and therefore sparse in
terms of the spatial sampling, these observations do provide profound
insights into the complexity of the shock acceleration phenomenon. In
astrophysical sites beyond the solar system, our role is passive, as
observers of signals from remote sites of acceleration.  Moreover, the information
on plasma properties is subject to a convolution with radiative
processes, complicated source morphology within our spatial resolution
scale, and propagational modification along the line of sight to sources. 

Observationally, radio, optical and X-ray telescopes have provided
groundbreaking insights into the shock acceleration phenomenon, due to
advances in the angular resolution and spectral sensitivity.  In
addition, the gamma-ray field is generating a greater understanding of
source energetics as we find that many non-thermal astronomical sources
emit most of their power in the gamma-rays. Progress on the theoretical
front builds on the observational advances, and has turned more to
computer simulations due to the dramatically enhanced speed of computers
over the last two decades.  Yet astrophysical code verification is an
increasingly salient issue as their complexity and computational demands
burgeon.  This provides a niche for laboratory plasma experiments that
are tailored for the problem of astrophysical particle acceleration. 
Supersonic flows can be generated in controlled environments, and work
along these lines focuses naturally on using lasers to mimic blast
waves and jets, supernovae and supernova remnants (e.g. Borovsky et al. 1984;
Drake et al. 1998; Shigemori, et al. 2000; Kang et al. 2001; Lebedev et al. 2002;
Woolsey, Courtois \& Dendy 2004), and probe hydrodynamic and magnetohydronamic
aspects.  The question of scalability of such laboratory findings to
astrophysical systems (e.g. Ryutov et al. 2001; see also the review of 
Remington, Drake \& Ryutov 2006) is obviously of central
importance. Here, an offering on some topical issues for particle
acceleration at astrophysical shocks is made, to provide a basis for the
community in high energy density plasma physics to help identify germane
astrophysical problems that might be well suited for interdisciplinary
investigation.

\section{Magnetic Field Enhancements in Shocks}
\label{sec:amplification}

One of the key properties of shock structure that is germane to the
acceleration of high energy cosmic rays is the strength of the magnetic
field \teq{B} near the shock.  In astrophysical shocks this cannot
really be measured directly, since there is generally a lack of viable
spectral line diagnostics: the Zeeman effect and cyclotron
emission/absorption features are generally broadened, small or
non-existent in diffuse, turbulent shock environs.  Normally, proximity
of an emission region to a stellar surface, such as in white dwarfs and
neutron stars is required to afford precise magnetic field measurements. 
In heliospheric shocks, magnetometer data discern the chaotic nature of
pre- and post-shock fields, and a prominent property appears to be (e.g.
Baring et al. 1997) a shock-induced compression of the field in the
downstream region that is fairly close to 
magnetohydrodynamic (MHD) determinations that are derived from momentum 
and energy flux conservation across the shock, i.e. the so-called
Rankine-Hugoniot conditions (e.g. Drury 1983; Jones \& Ellison 1991). 
If the interaction of charged particles with shock-associated field
turbulence is {\bf gyroresonant} at the Doppler-shifted cyclotron
frequency (e.g., see Melrose 1980), as is expected for Alfv\'en and
whistler modes, then the acceleration timescale \teq{\tau} naturally
scales as the gyroperiod (i.e., \teq{\tau\sim 1/\nu_g\propto 1/B}; see
Forman, Jokipii \& Owens 1974; Drury 1983) and the corresponding
diffusive lengthscale is comparable to the Larmor radius \teq{r_g} 
(\teq{\propto 1/B}). Hence the magnitude \teq{B} establishes the temporal, 
spatial and energy scales of acceleration at a shock, and so is a critical 
parameter for the energization process.

Since the general paradigm of galactic cosmic rays (CRs) invokes
supernova remnants (SNRs) as the sites for their production (see Drury
1983 for a review), knowledge of the field strength in proximity of
their shocks is vital.  Directional information can be obtained on
fairly large spatial scales via radio synchrotron polarization data 
(e.g. see Rosenberg 1970; Downs \& Thompson 1972; Anderson, 
Keohane \& Rudnick 1995, for Cassiopeia A), but values of 
\teq{B=\vert \hbox{\bf B} \vert} are not
forthcoming. Estimates for the field strength can be inferred by
modeling the continuum flux level in a given wavelength band, but these
are subject to a number of assumptions about the medium, for example the
mean density \teq{n_e} and the Lorentz factors \teq{\gamma_e} of the
radiating electrons. The understanding of the character of shells and
interiors of SNRs has recently been advanced by groundbreaking
observations with the \Chandra\ X-ray Observatory, enabled by its
impressive angular resolution coupled with its spectral capabilities. 
Of particular interest is the observation of extremely narrow
non-thermal (1.2 -- 2.0 keV) X-ray spatial profiles in selected remnants
(see Long et al. 2003 and Bamba et al. 2003 for the northeast limb of
SN1006; Vink \& Laming 2003 for Cas A; for theoretical modeling see
Ellison \& Cassam-Chena\"i 2005, and V\"olk et al. 2005), typically less
than 5--10 arcsec across.  Upstream of these shell shocks, the X-ray
emission, which is thought to be synchrotron in origin, drops to
effectively zero.  These strong brightness contrasts between the shell,
and the outer, upstream zones correspond to flux ratios exceeding
\teq{R\gtrsim 50}.  The narrowness of profiles along image
scans argues for the shocks being aligned perpendicular to the sky,
i.e., offering no projectional smearing in the images.  Note also that
the surface brightness angular profiles in SN1006 and Cas A are much broader for
the thermal X-rays (0.5--0.8 keV) and the radio synchrotron than for the
non-thermal X-rays.

If the synchrotron mechanism is indeed responsible for non-thermal
\Chandra\ emission, the electrons contributing to the \Chandra\
signal are probably in a strongly-cooling regime: see Baring et al.
(1999) for a comprehensive discussion of SNR cooling parameter space.
Since the synchrotron cooling rate for an electron scales as
\teq{\gamma_e^2B^2}, then the flux ratio \teq{R} is
approximately a measure of the ratio of \teq{B^2} downstream (d) to
upstream (u).  The observed lower bounds to \teq{R} considerably exceed values
\teq{R\lesssim 16} expected for magnetohydrodynamic compression at the 
shocked shell; at a plane-parallel shock with {\bf B} along the shock normal
there is no field compression, while in a strong (i.e. high sonic Mach
number) perpendicular shock with {\bf B} in the shock plane, \teq{B_d
/B_u\sim 4}.  Hence, the pronounced brightness contrast is taken
as strong evidence of {\bf magnetic field amplification} in the shock
precursor/ramp upstream. Higher fields are obviously advantageous to
cosmic ray production issues. Historically-accepted values of \teq{B\sim
1}--\teq{10\mu}Gauss (i.e. 0.1--1 nanoTesla) are somewhat too small to
permit acceleration in SNR shocks of ages around \teq{10^3}--\teq{10^4}
years right up to the {\it cosmic ray knee} at \teq{\sim 3\times
10^{15}}eV (e.g. see Lagage \& Cesarsky 1983).  This problem has spawned
the suggestion (Jokipii 1987) that relatively ineffective diffusive transport of particles
orthogonal to the mean field direction in
quasi-perpendicular regions of SNR shocks can speed up acceleration of
ions to higher
energies, helping access the knee.  Yet, this enhanced rapidity is
accompanied by reduced efficiency of cosmic ray injection from thermal
energies (Ellison, Baring \& Jones 1995).  Hence, truly larger fields
provide a cleaner path for acceleration in remnants to reach the cosmic
ray knee (e.g. Kirk \& Dendy 2001).

The X-ray observational developments have been accompanied by
theoretical proposals of magnetic field amplification in the upstream
shock precursor.  Most notable has been the work of Lucek \& Bell
(2000), and subsequent papers such as Bell (2004), Amati \& Blasi (2006),
and Vladimirov, Ellison \& Bykov (2006).  The idea of Lucek \&
Bell is that high energy cosmic rays in strong shocks could amplify
\teq{B} when streaming upstream, adiabatically transferring energy
to the turbulent field by pushing against it, simultaneously decelerating the upstream
flow.  If this process is
efficient, the rate of work done on the upstream Alfv\'en turbulence of
energy density \teq{\Ualf} naturally scales roughly with the CR pressure
gradient: \teq{d\Ualf /dt = \valf \vert \nabla P_{\rm CR}\vert}. Here
\teq{\valf =B/\sqrt{4\pi\rho}} is the Alfv\'en speed, 
and \teq{P_{\rm CR}} is the cosmic
ray pressure. The associated field amplification should then scale as
\teq{(\delta B/B)^2 \sim \Machalf P_{\rm CR}/\rho u_u^2} in an upstream
flow of speed \teq{u_u} and mass density \teq{\rho}; this then becomes
very effective for high Alfv\'enic Mach number (i.e. \teq{\Machalf
\equiv u_u/\valf\gg 1}), strong shocks that generate large cosmic ray
pressures. While this hypothesis is reasonable, demonstrating it is
non-trivial.  Various MHD-type simulations have been employed by Bell,
such as in Bell (2004), where large-scale currents are used to drive
instabilities that amplify the upstream field.  The persistence of
currents on large scales is unclear, particularly due to the action of
Debye screening.   Moreover, self-consistent physical connection between
the cosmic rays of large Larmor radii and the field turbulence of much
shorter wavelengths is extremely difficult to explore with MHD or plasma
simulations, due to the wide disparity in spatial scales involved.  This
is an issue also for a growing number of particle-in-cell (PIC)
simulations (e.g. Silva et al. 2003; Hededal et al. 2004; Nishikawa et
al. 2005; see Section~\ref{sec:relshock} below) used to explore field 
enhancement via the Weibel instability in relativistic shocks; such 
developments are not that salient for the problem
of amplifying Alfv\'en turbulence in non-relativistic shocks, and mostly
probe the inertial scales of thermal ions and electrons defined by their
plasma frequencies.

\section{Non-linear Feedback Between the Acceleration and the Hydrodynamics}
 \label{sec:non-linear}

Non-relativistic collisionless shocks can be highly efficient
accelerators, placing 10--50\% of the bulk flow kinetic energy into
non-thermal particles.  Evidence from theory, computer simulations, and
spacecraft observations supports this conclusion; in particular, see
Ellison, M\"obius, \& Paschmann (1990), for a study of the Earth's bow
shock, and Drury (1983), Blandford \& Eichler (1987), and Jones \&
Ellison (1991) for reviews.  With such efficiencies, the accelerated
particles acquire a sizable fraction of the total energy budget,
influencing the shock hydrodynamics, and therefore also the fraction of
energy going into accelerated particles, in a non-linear manner.  The
modified flow velocity spatial profile in the shock deviates from the
familiar step-function form in test-particle acceleration scenarios,
with the energetic particles pushing against the upstream flow and
decelerating it far ahead of the shock discontinuity.  Accordingly an
upstream shock precursor forms, with declining flow velocity as the
shock is approached. This structure alters the shape of the energetic
particle distribution from a power-law in momentum (e.g., Eichler 1984;
Ellison \& Eichler 1984; Ellison, Baring \& Jones 1996; Berezhko et al.
1996; Malkov 1997; Blasi 2002), the canonical test-particle form where
the diffusively-accelerated particle distribution samples no spatial or
momentum scale.  The index \teq{\sigma =(r+2)/(r-1)} of this power-law
\teq{dn/dp\propto p^{-\sigma}} is purely a function of the compression ratio
\teq{r=u_u/u_d} of upstream (\teq{u_u}) to downstream (\teq{u_d}) flow
speed components normal to the shock in the shock rest frame (e.g. see
Drury 1983; Jones \& Ellison 1991), and is independent of the magnetic
field orientation or the nature and magnitude of the turbulence
effecting diffusive transport in the shock neighborhood.

The spatial variation of the upstream flow in strong shocks that are efficient 
accelerators eliminates the scale independence.  Since the highest
energy particles have greater diffusive mean free paths \teq{\lambda}
(generally true for gyroresonant interactions with MHD turbulence, and
certainly so near the Bohm diffusion limit \teq{\lambda\sim r_g}), they
diffuse farther into the upstream shock precursor against the convective
power of the flow, and therefore sample greater effective velocity
compression ratios \teq{r}.  Accordingly, they have a flatter
distribution, yielding a distinctive concavity to the overall particle
spectrum, i.e. \teq{\sigma} is now a declining function of momentum
\teq{p}.  These departures from power-law behavior amplify the energy
placed in the particles with the greatest momenta, which in turn feeds
back into the shock hydrodynamics that modify the spatial flow velocity
profile. Traveling discontinuities possessing this complex feedback
are termed {\bf non-linear shocks}, the non-linear label being ascribed
to the interplay between the macroscopic dynamics and the microscopic
acceleration process.  Clearly, the possible magnetic field amplification 
in the upstream precursor that was discussed in Section~\ref{sec:amplification} 
contributes to the overall dynamics/energy budget of the magnetohydrodynamic 
flow, and so intimately influences this non-linear aspect of astrophysical shocks.

The deviations from power-law distributions obviously impact
the radiation signatures produced by these particles, with alterations
in the fluxes expected in X-ray and TeV gamma-ray bands in remnants,
differing by as much as factors of 3--10 from traditional test-particle
predictions (e.g.,  see Baring, et al. 1999; Ellison, Slane \& Gaensler
2001; Berezhko et al. 2002; Baring, Ellison \& Slane 2005). 
Conclusively confirming the existence of this non-linear spectral
concavity is a major goal that is inherently difficult, since it demands
broad, multi-wavelength spectral coverage.  There is a limited
suggestion of concavity in radio data for Tycho's and Kepler's SNRs 
(Reynolds \& Ellison 1992), and in a multi-wavelength modeling of SN 1006 (Allen, Houck 
\& Sturner 2004; see also Jones et al. 2003 for inferences from radio and
infra-red data from Cassiopeia A), 
but this task really looks ahead to the launch of the GLAST
gamma-ray mission in late 2007, when, in conjunction with ground-based
Atmospheric \v{C}erenkov Telescopes probing the TeV band, it may prove 
possible to determine gamma-ray spectra from SNRs spanning over 
3 decades in energy.

In the meantime, an interesting astrophysical manifestation of these
non-linear effects has been offered by SNR observations by the \Chandra\
X-ray Observatory, looking instead at the thermal populations.
Inferences of ion temperatures in remnant shocks can be made using
proper motion studies, or more direct spectroscopic methods (e.g.
Ghavamian et al. 2003).  For the remnant 1E 0101.2-7129, Hughes et al.
(2000) used a combination of ROSAT and Chandra data spanning a decade to
deduce an expansion speed. Electron temperatures \teq{T_e} are
determined by line diagnostics, via both the widths, and the relative
strengths for different ionized species.  From these two ingredients,
Hughes et al. (2000) observed that, in selected portions of the SNR shell, 
\teq{3kT_e/2\ll 3kT_p/2\sim m_p (3u_u/4)^2 /2}.  Therefore, the electrons 
were considerably cooler than would correspond to equipartition 
with thermal protons heated in a strong shock
with an upstream flow speed of \teq{u_u}: the
thermal heating is assumed comparable to the kinematic velocity
differential \teq{u_u-u_d\approx 3u_u/4}.  The same inference was made
by Decourchelle et al. (2000) for Kepler's remnant, and by Hwang et al.
(2002) for Tycho's SNR.  This property of comparatively cooler electrons
may be indicative of them radiating very efficiently radiating.  Or
it may suggest that the protons are cooler (i.e.  \teq{3kT_p/2\ll m_p
(3u_u/4)^2 /2}) than is widely assumed in the test-particle theory, the conclusion 
drawn by Hughes et al. (2000) and Decourchelle et al. (2000).  This
effect is naturally expected in the non-linear shock
acceleration scenario:  as the highest energy particles tap significant
fractions of the total available energy, they force a reduction in the
thermal gas temperatures. Such feedback can profoundly
influence shock layer thermalization, inducing significant
interplay with electrostatic equilibration between low energy electrons
and ions, an issue addressed in these proceedings by Baring \& Summerlin
(2006). Note that non-linear modifications may vary strongly around the
shocked shell of an SNR, since the obliquity angle \teq{\ThetaBnone} of
{\bf B} to the shock normal varies considerably between different rim
locales.

\newpage

\section{The Character of Relativistic Shocks}
 \label{sec:relshock}

Relativistic shocks, for which the upstream flow 
Lorentz factor \teq{\gamma_u=1/\sqrt{1- (u_u/c)^2}} considerably exceeds unity,
are less well researched than their non-relativistic
counterparts, not in small part due to their greater cosmic remoteness:
they predominantly arise in extragalactic locales like jets in active galaxies,
and gamma-ray bursts.  Yet, because of such associations, they are 
now quite topical.  Diffusive test-particle acceleration theory in parallel, 
(i.e., \teq{\ThetaBnone =0^{\circ}}) relativistic shocks identifies two notable 
properties in such systems:
(i) particles receive a large energy kick \teq{\Delta E \sim \gamma_u m c^2} in 
    their first shock crossing (e.g., Vietri 1995), but receive much smaller energy 
    boosts for subsequent crossing cycles (factors of around two: e.g., 
    Gallant \& Achterberg 1999; Baring 1999);
(ii) a so-called `universal' spectral index, \teq{\sigma \sim 2.23} exists 
    in the two limits of \teq{\gamma_u \gg 1} and small angle scattering, 
    i.e., \teq{\delta\theta \ll 1/\gamma_u} (e.g., Kirk et al. 2000; see also
    Bednarz \& Ostrowski 1998; Baring 1999; Ellison and Double 2004). Here, 
    \teq{\delta\theta} is the average angle a particle's momentum vector deviates 
    in a scattering event, i.e. an interaction with magnetic turbulence.

These characteristics are modified in parallel, mildly relativistic
shocks with \teq{\gamma_u\sim 1}. In such shocks, the distribution
\teq{dn/dp} remains a power-law (scale-independence persists), but
hardens (\teq{\sigma} decreases) as either \teq{\gamma_u} drops, or the
scattering angle, \teq{\delta\theta}, increases (e.g., Ellison, Jones \&
Reynolds 1990; Baring 1999; Ellison \& Double 2004; Baring 2004), even
if the compression ration \teq{r=u_u/u_d} is held constant (it usually
increases with declining \teq{\gamma_u} due to a hardening of the
J\"uttner-Synge equation of state).  These effects are consequences of
large kinematic energy kicks particles receive when scattered in the
upstream region after transits from downstream of the shock.  It is
particularly interesting that when scattering conditions deviate from
fine pitch-angle-scattering regimes with \teq{\delta\theta \ll
1/\gamma_u}, the power-law index is dependent on \teq{\delta\theta},
with a continuum of spectral indices being possible (Ellison \& Double
2004; Baring 2004).  Then the nature of the turbulence is extremely
influential on the acceleration outcome, so that understanding the
turbulence is of paramount importance.  This sensitivity of \teq{\sigma}
to the field fluctuations when \teq{\delta\theta \gtrsim 1/\gamma_u}, a
large angle scattering domain, contrasts the canonical nature of
\teq{\sigma} in non-relativistic shocks mentioned above.

In jets and gamma-ray bursts, ultra-relativistic shocks are typically
highly oblique due to the Lorentz transformation of ambient, upstream
magnetic fields to the shock rest frame.  This introduces an added
dimension of variation, with increasing \teq{\ThetaBnone} dramatically
steepening the power-law, i.e. increasing \teq{\sigma}. This is
naturally expected since such systems are highly superluminal, that is,
there exists no de Hoffman-Teller (1950) shock rest frame where the flow 
velocities are everywhere parallel to the mean magnetic field (which
would correspond to large 
scale electric fields being zero everywhere).  Therefore, relativistic shocks are much
less efficient accelerators because particles convect more rapidly away
downstream from the shock (e.g. Begelman \& Kirk 1990). In oblique,
relativistic shocks, \teq{\sigma}, and indeed the efficiency of
injection from the thermal particle population, also depend on the
ability of turbulence to transport particles perpendicular to the mean
downstream field direction (Ellison \& Double 2004; Niemiec \& Ostrowski
2004). This perpendicular transport couples directly to the magnitude
\teq{(\delta B/B)^2} and power spectrum of field fluctuations, i.e. the
strength of the scattering.  Steep spectra (\teq{\sigma\gtrsim 4})
result unless the ratio of the diffusive mean free paths perpendicular
to and parallel to {\bf B} is comparable to unity, which defines the
{\it Bohm diffusion} regime.  In summation, for relativistic shocks, the
spectral index is sensitive to the obliquity \teq{\ThetaBnone} of the
shock, the nature of the scattering, and the strength of the turbulence
or anisotropy of the diffusion.  These properties are reviewed in Baring
(2004).

Observational vindication of these theoretical predictions is clearly
mandated. This is not readily forthcoming, since the only accessible
information involves a convolution of shock acceleration and radiation
physics. Yet, it is clear, for example in gamma-ray bursts (GRBs), that
data taken from the EGRET experiment on the Compton Gamma-Ray
Observatory (CGRO) suggest a broad range of spectral indices (Dingus
1995) for the half dozen or so bursts seen at high energies.  This
population characteristic is commensurate with the expected
non-universality of \teq{\sigma} just discussed. Yet it is important to
emphasize that the power-law index is not the only acceleration
characteristic germane to the GRB problem: the shapes of the particle
distributions at thermal and slightly suprathermal energies are also
pertinent.  This energy domain samples particle injection or
dissipational heating in the shock layer, and is readily probed for
electrons by the spectrum of prompt GRB emission by the BATSE instrument
on CGRO.  Tavani (1996) obtained impressive spectral fits to several
bright BATSE bursts using a phenomenological electron distribution and
the synchrotron emission mechanism. While there are issues with fitting
low energy (i.e. \teq{\lesssim 100}keV) spectra in about 1/3 of bursts
(e.g. Preece et al. 1998) in the synchrotron model, this radiative
mechanism still remains the most popular candidate today for prompt
burst signals.

Tavani's work was extended recently by Baring \& Braby (2004), who
provided additional perspectives, using acceleration theory to underpin
a program of spectral fitting of GRB emission using a sum of thermal and
non-thermal electron populations. These fits demanded that the
preponderance of electrons that are responsible for the prompt emission
constitute an {\bf intrinsically non-thermal population}.  That is, the contribution
to the overall electron distribution that comes from a Maxwell-Boltzmann
distribution is completely dominated by a non-thermal contribution that, to
first order, can be approximated by a power-law in energy truncated at 
some minimum electron Lorentz factor.  This requirement of non-thermal 
dominance strongly contrasts particle distributions obtained from
acceleration simulations, as is evident in a host of the references
cited on acceleration theory above: the non-thermal particles are drawn
directly from a thermal gas, a virtually ubiquitous phenomenon.   This
conflict poses a problem for acceleration scenarios unless (i) radiative
efficiencies for electrons in GRBs only become significant at highly
superthermal energies, or (ii) shock layer dissipation in relativistic
systems can suppress thermalization of electrons. A potential resolution
to this dilemma along the lines of option (i) is that strong radiative
self-absorption could be acting, in which case the BATSE spectral probe
is not actually sampling the thermal electrons.  It is also possible
that other radiation mechanisms such as Compton scattering, pitch-angle
synchrotron, or jitter radiation may prove more germane.  Discerning the
radiation mechanism(s) operating in bursts is a foremost goal of future
research, and will be facilitated by the GLAST mission, with its
good sensitivity in the 5 keV -- 300 GeV band, in conjunction
with NASA's current GRB flagship venture, Swift.

Option (ii) is a conjecture that has no definitive simulational evidence
to support it at present. The most comprehensive way to study
dissipation and wave generation in collisionless shocks is with PIC
simulations, where particle motion and field fluctuations are obtained
as solutions of the Newton-Lorentz and Maxwell's equations. Rich in
their turbulence information, these have been used extensively in
non-relativistic, heliospheric shock applications, and more recently,
relativistic PIC codes have blossomed to model shocks in various
astrophysical systems.  PIC simulation research has largely, but not
exclusively, focused on perpendicular shocks, first with Gallant et
al.(1992), Hoshino et al. (1992), and then Smolsky \& Usov (1996),
Shimada \& Hoshino (2000), Silva et al. (2003), Nishikawa et al. (2003,
2005), Spitkovsky \& Arons (2004), Hededal et al. (2004), Liang \&
Nishimura (2004), Medvedev et al. (2005) and Hededal \& Nishikawa
(2005). These works have explored pair shocks, ion-doped shocks,
Poynting flux-dominated outflows, and low-field systems with dissipation
driven by the Weibel instability, in applications such as GRBs and
pulsar wind termination shocks. PIC simulations are dynamic in nature,
and rarely achieve a time-asymptotic state. Even in the minority of
cases where there is some evidence of acceleration beyond true
thermalization, none of these works has demonstrated the establishment
of an extended power-law that is required in modeling emission from GRBs
and active galactic nuclei. This is perhaps due to the severely
restricted spatial and temporal scales of the simulations, imposed by
their intensive CPU and memory requirements; these limit the modeling of
realistic electron-to-proton mass ratios, full exploration of
three-dimensional shock physics such as diffusive transport, and
addressing the wide range of particle momenta encountered in the shock
acceleration process.  In particular, it is difficult to establish a broad inertial
range for cascading MHD turbulence when the maximum spatial scales 
in the simulation are not orders of magnitude larger than the principal
ion inertial scales.  Definitively observing the injection of electrons
from a thermal population to establish a truly non-thermal distribution
remains a pressing goal of plasma simulations, both for relativistic
shocks and, as has been understood for more than two decades, also their
non-relativistic cousins.

\section{Conclusion}

This review is by no means a complete presentation of the topical issues
for the shock acceleration problem, but it does offer a fair sampling
suitable for motivating interdisciplinary activity.  It is clear that
several issues could benefit substantially from input from laboratory
experimentation on the high energy density physics/astrophysics
interface.  One key question is whether or not ambient magnetic fields
are amplified by both non-relativistic and relativistic shocks beyond
standard MHD expectations.  If so, is the amplification electrostatic in
origin, or is it connected to energetic particles accelerated by the
shock?  It would be important to discern whether there are differences
between high and low Alfv\'enic Mach number systems, i.e. what role the
ambient magnetic field plays in controlling the outcome. Another question
concerns whether or not suprathermal electrons and ions can actually be
seen, and whether one can identify their origin. It is salient to
ascertain if they are diffuse in nature, or if they form coherent beams,
both of which are seen at traveling shocks embedded in the solar wind.
Also, if acceleration is observed, then identifying the role the highest
energy particles have in modifying the shock hydrodynamics and the
thermal structure of the shock layer would help solve an outstanding
problem that has long been a principal goal within the cosmic ray
community.  Finally, specifically concerning relativistic systems, it
would be desirable to elucidate how the distributions of any accelerated
particles seen depend on external quantities such as the field obliquity
and speed of the shock, whether thermal electrons can be suppressed
relative to accelerated ones, and if there is an identifiable connection
with the field turbulence near the shock. These are demanding goals, yet
terrestrial experiments are very useful for probing global aspects of
shock problems, and in particular for extracting insights into
hydrodynamic and MHD behavior.  In order to make progress, it is
essential to prepare an experimental setup that is as tenuous as
possible, to mimic the collisionless (in the Spitzer sense) shock
environments offered throughout the cosmos. At this juncture, exciting
prospects are on the horizon for this interdisciplinary forum, with
contributions to be found in the laboratory, in computer simulations,
and in astronomical observations, all of which can benefit from
cross-fertilization with each other.

\vskip 6pt
\noindent {\bf References:}
\vskip 2pt
\def\ref{\noindent \hangafter=1 \hangindent=0.7 true cm}

\def\mn{M.N.R.A.S.}
\def\aas{{Astron. Astrophys.}}
\def\aassupp{{Astron. Astrophys. Supp.}}
\def\apss{{Astr. Space Sci.}}
\def\aj{AJ}
\def\nat{Nature}
\def\aaps{{Astron. \& Astr. Supp.}}
\def\xviiicrc{{17th Internat. Cosmic Ray Conf. Papers}}
\def\xviiiicrc{{18th Internat. Cosmic Ray Conf. Papers}}
\def\xixicrc{{19th Internat. Cosmic Ray Conf. Papers}}
\def\xxicrc{{20th Internat. Cosmic Ray Conf. Papers}}
\def\xxiicrc{{21st Internat. Cosmic Ray Conf. Papers}}
\def\apj{{ApJ}}
\def\aa{{A\&A}}
\def\apjs{{ApJS}}
\def\sp{{Solar Phys.}}
\def\jgr{{J. Geophys. Res.}}
\def\grl{{Geophys. Res. Lett.}}
\def\jphysb{{J. Phys. B}}
\def\ssr{{Space Science Rev.}}
\def\araa{{Ann. Rev. Astron. Astrophys.}}
\def\nature{{Nature}}
\def\asr{{Adv. Space. Res.}}
\def\prc{{Phys. Rev. C}}
\def\prd{{Phys. Rev. D}}
\def\pr{{Phys. Rev.}}
\def\rgph{{Rev. Geophys.}}
\def\cjp{{Canadian J. Physics}}
\def\np{{Nuclear Physics}}
\def\sturrock{{Solar Flares}, ed. P. A. Sturrock,
       (Boulder:Colorado Associated University Press)}
\def\accaip{{Particle Acceleration Mechanisms in 
Astrophysics}, ed. J.~Arons, C.~Max, C.~McKee (New York: AIP)}
\def\ljaip{{Gamma Ray Transients and Related Astrophysical Phenomena,
            } eds. R. E. Lingenfelter, H. S. Hudson, and D. M. Worrall
            (New York:AIP)}
\def\gsfcaip{{Positron Electron Pairs in Astrophysics},
    eds. M. L. Burns, A. K. Harding, and R. Ramaty (New York:AIP)}
\def\washaip{{Nuclear Spectroscopy of Astrophysical
    Sources}, eds. N. Gehrels and G. H. Share, (New York:AIP)}
\def\minaip{{Cosmic Abundances of Matter}, ed. C. J. Waddington,
(New York:AIP)}
\def\gcacta{{Geochim. Cosmochim. Acta}}
\def\mnras{{M.N.R.A.S.}}
\def\ref{\noindent\hangafter=1\hangindent=1truecm}

\small
{\baselineskip 12.6pt

\ref
Allen, G.~E., Houck, J.~C. \& Sturner, S.~J. 2004, 
in on-line proceedings of the {\it X-ray and Radio Connections}
     workshop, Sante Fe, New Mexico. {\tt [http://www.aoc.nrao.edu/events/xraydio]}

\ref
Amati, E. \& Blasi, P. 2006, \mnras,\vol{371}{1251}

\ref
Anderson, M.~C., Keohane, J.~W. \& Rudnick, L. 1995, \apj,\vol{441}{300}

\ref
Bamba, A., Yamazaki, R., Ueno, M. \& Koyama, K. 2003, \apj,\vol{589}{827}

\newpage

\ref
Baring, M. G.  1999, 
in \it Proc. of the 26th International Cosmic Ray Conference, Vol. IV \rm ,
     p.~5, held in Salt Lake City, Utah,    {\tt [astro-ph/9910128]}.

\ref
Baring, M.~G. 2004, Nuclear Physics B, Proc. Supp., \vol{136}{198}
   
\ref
Baring, M.~G. \& Braby, M.~L. 2004, \apj\vol{613}{460}

\ref
Baring, M.~G., Ellison, D.~C., Reynolds, S.~P., Grenier, I.~A., \& Goret, P.
1999, \apj,\vol{513}{311}

\ref
Baring, M.~G., Ellison, D.~C., \& Slane, P.~O. 2005,  \asr,\vol{35}{1041}

\ref
Baring, M.~G.,  Ogilvie, K.~W., Ellison, D.~C. \& Forsyth, R.~J. 1997,
	\apj,\vol{476}{889}

\ref
Baring, M.~G. \& Summerlin, E.~J. 2006, \apss, {\it in press}.  {\tt [astro-ph/0609407]}

\ref
Bednarz, J. \& Ostrowski, M. 1998, \prl,\vol{80}{3911}

\ref
Begelman, M.~C. \& Kirk, J.~G. 1990, \apj,\vol{353}{66}

\ref
Bell, A.~R., 2004, \mnras,\vol{353}{550}

\ref
Berezhko, E.~G., Ksenofontov, L.~T. \& V\"olk, H.~J. 2002, \apj,\vol{395}{943}
   
\ref
Berezhko, E.~G., Yelshin, V.~K. \& Ksenofontov, L.~T. 1996,
	JETP,\vol{82(1)}{1} 

\ref
Blandford, R.~D. \&  Eichler, D. 1987, Phys. Rep.,\vol{154}{1} 

\ref
Blasi, P. 2002, \app,\vol{16}{429}

\ref
Borovsky, J.~E., Pongratz, M.~B., Roussel-Dupre, R.~A. \& Tan, T.-H. 1984, \apj,\vol{280}{802}

\ref
Decourchelle, A., Ellison, D.~C. \& Ballet, J. 2000, \apj,\vol{543}{L57} 

\ref
de Hoffman, F. \& Teller, E. 1950, \prd,\vol{80}{692}

\ref
Dingus, B.~L. 1995, \apss,\vol{231}{187}

\ref
Downs, G.~S. \& Thompson, A.~R. 1972, \aj,\vol{77}{120}
 
\ref
Drake, R.~P., et al. 1998, \apjl,\vol{500}{L157}

\ref
Drury, L.~O'C. 1983, Rep. Prog. Phys.,\vol{46}{973}

\ref
Eichler, D. 1984, \apj,\vol{277}{429} 

\ref
Ellison, D.~C., Baring, M.~G. \& Jones, F.~C. 1995, \apj,\vol{453}{873}

\ref
Ellison, D.~C., Baring, M.~G. \& Jones, F.~C. 1996, \apj,\vol{473}{1029}

\ref
Ellison, D.~C. \& Cassam-Chena\"i, G. 2005, \apj,\vol{632}{920}

\ref
Ellison, D.~C. \& Double, G.~P. 2004, \app,\vol{22}{323}

\ref
Ellison, D.~C. \& Eichler, D. 1984, \apj,\vol{286}{691}

\ref
Ellison, D.~C., Jones, F.~C., \& Reynolds, S.~P. 1990, \apj,\vol{360}{702}

\ref 
Ellison, D.~C., M\"obius, E. \& Paschmann, G. 1990 \apj,\vol{352}{376}

\ref
Ellison, D.~C., Slane, P. \& Gaensler, B.~M. 2001 \apj,\vol{563}{191}

\ref
Forman, M.~A., Jokipii, J.~R. \& Owens, A.~J. 1974 \apj,\vol{192}{535}

\ref
Gallant, Y.~A. \& Achterberg, A. 1999, \mnras,\vol{305}{L6}

\ref
Gallant, Y.~A.,  Hoshino, M.,  Langdon, A.~B.,  Arons, J., \& Max, C.E. 
	1992 \apj,\vol{391}{73}
	
\ref
Ghavamian, P., Rakowski, C.~E., Hughes, J.~P. \& Williams, T.~B. 2003, \apj,\vol{590}{833}

\ref
Hededal,ÊC.~B., Haugbolle,ÊT., Frederiksen,ÊJ.~T. \& Nordlund,ÊA 2004,
\apj,\vol{617}{L107}

\ref
Hededal, C.~B. \& Nishikawa, K.-I. 2005, \apj,\vol{623}{L89}

\ref
Hoshino, M., Arons, J.,  Gallant, Y.~A. \& Langdon, A.~B. 1992,
	\apj,\vol{390}{454}

\ref
Hughes, J.~P., Rakowski, C.~E., \& Decourchelle, A. 2000, \apj,\vol{543}{L61} 

\ref
Hwang, U., et. al. 2002, \apj,\vol{581}{110}  

\ref
Jokipii, J.~R. 1987, \apj,\vol{313}{842}

\ref
Jones, F.~C. \& Ellison, D.~C. 1991, \ssr,\vol{58}{259}

\ref
Jones, T.~J., Rudnick, L., DeLaney, T. \& Bowden, J. 2003, \apj,\vol{587}{338}

\ref
Kang, Y.-G., et al. 2001, Plasma Phys. Rep., \vol{27}{843}

\ref
Kirk, J.~G. \& Dendy, R.~O. 2001, J. Phys. G.,\vol{27}{1589}

\ref
Kirk, J.~G., Guthmann, A.~W., Gallant, Y.~A., Achterberg, A. 2000,
	\apj,\vol{542}{235}

\ref
Lagage, P.~O. \& Cesarsky, C.~J. 1983, \aap,\vol{125}{249}

\ref
Lebedev, S.~V., et al. 2002, \apj,\vol{564}{113}

\ref
Liang, E.~P. \& Nishimura, K. 2004, \prl,\vol{92}{5005}

\ref
Long, K.~S., et al. 2003, \apj,\vol{586}{1162}

\ref
Lucek, S.~G. \&  Bell, A.~R. 2000, \mnras,\vol{314}{65}

\ref
Malkov, M.~A.  1997, \apj,\vol{485}{638}

\ref
Medvedev, M.~V., et al. 2005, \apj,\vol{618}{L75}

\ref 
Melrose, D. B. 1980, Plasma Astrophysics, Vols. I \& II 
   (Gordon \& Breach, New York)

\ref
Niemiec, J., \& Ostrowski, M. 2004, \apj,\vol{610}{851}

\ref
Nishikawa, K.-I., et al. 2003, \apj,\vol{595}{555}

\ref
Nishikawa, K.-I., et al. 2005, \apj,\vol{622}{927}

\ref
Preece, R.~D., et al. 1998, \apj,\vol{506}{L23}

\ref
Remington, B.~A., Drake, R.~P. \& Ryutov, D.~D. 2006, Rev. Mod. Phys. \vol{78}{755}

\ref
Reynolds, S.~P. \& Ellison, D.~C. 1992, \apj,\vol{399}{L75}

\ref
Rosenberg, I. 1970, \mnras,\vol{151}{109}

\ref
Ryutov, D.~D., Remington, B.~A., Robey, H.~F., Drake, R.~P. 2001, Phys. Plasmas, \vol{8}{1804}

\ref
Shigemori, K., et al. 2000, \apj,\vol{533}{L159}

\ref
Shimada, N. \& Hoshino, M. 2000, \apj,\vol{543}{L67}
   
\ref
Silva, L.~O., et al. 2003, \apj,\vol{596}{L121}

\ref
Smolsky, M.~V. \& Usov, V.~V. 1996, \apj,\vol{461}{858}

\ref
Spitkovsky, A. \& Arons, J. 2004, \apj,\vol{603}{669}

\ref
Tavani, M. 1996, \prl,\vol{76}{3478}

\ref
Vietri, M. 1995, \apj,\vol{453}{883}

\ref
Vink, J. \& Laming, J.~M. 2003, \apj,\vol{584}{758} 

\ref
Vladimirov, A., Ellison, D.~C. \& Bykov, A. 2006, \apj, {\it in press}. {\tt [astro-ph/0606433]}

\ref
V\"olk, H.~J., Berezhko, E.~G. \& Ksenofontov, L.~T. 2005, \aap,\vol{433}{973}

\ref
Woolsey, N.~C., Courtois, C. \& Dendy, R.~O. 2004, Plasma Phys. Cont. Fus., \vol{46}{B397}

}

\end{document}